\documentstyle[aps,prl,multicol,epsf]{revtex}
\def\Bildeins{\epsfbox{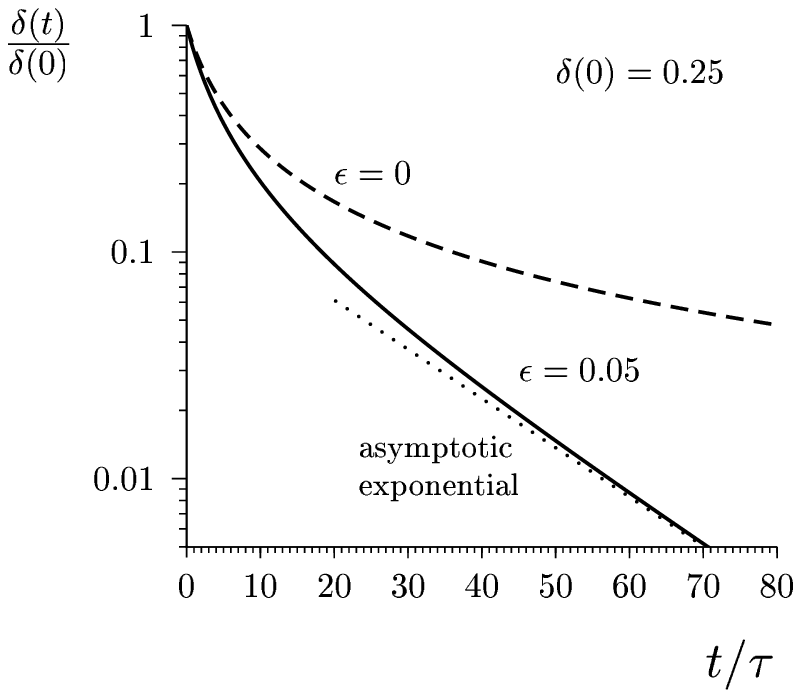}}
\def\ad{c_{\downarrow}}      \def\au{c_{\uparrow}}
\def\adp{c_{\downarrow p}}   \def\aup{c_{\uparrow p}}
\def\vec{\bf}
\def\sig{\mbox{\boldmath $\sigma$}}
\def\lam{\mbox{\boldmath $\lambda$}}
\def\delt{\mbox{\boldmath $\delta$}}
\def\be{\begin{equation}}          \def\ee{\end{equation}}
\def\bea{\begin{eqnarray}}         \def\eea{\end{eqnarray}}
\begin{document}

\title{Spin relaxation in quantum Hall systems}

\author{W. Apel$^{1}$ and Yu.~A. Bychkov$^{2,1}$ }

\address{$^{1}$ Physikalisch-Technische Bundesanstalt,
Bundesallee 100, 38116 Braunschweig, Germany.}

\address{$^{2}$ L.D. Landau Institute for Theoretical Physics,
ul.Kosigina, 2, Moscow, Russia.}

\date{\today}

\maketitle

\begin{abstract}
We study the spin relaxation in an interacting two--dimensional electron gas
in a strong magnetic field for the case that the electron density is
close to filling just one Landau sub--level of one spin projection,
i.e., for filling factor $\nu \simeq 1$.
Assuming the relaxation to be caused by scattering with phonons, we
derive the kinetic equations for the electron's spin--density
which replace the Bloch equations in our case.
These equations are non--linear and their solution depends crucially on
the filling factor $\nu$ and on the temperature $T$ of the phonon bath.
In the limit of $T=0$ and $\nu = 1$, the solution relaxes asymptotically
with a power law inversely proportional to time,
instead of following the conventional exponential behavior.
\end{abstract}

\begin{multicols}{2}

\vspace{3ex}

Experiments in two--dimensional (2-d) electronic structures in a
strong magnetic field reveal a rich variety of features.
Recently, many experiments have focussed on properties related to the
electrons' spins
\cite{NHKW88,UNHF90,BDPWT95,SEPW95,AGB96,MPPHEHP96,KKE97,LNMUPHF97}.
Optical measurements of the magnetization and transport measurements
of the activation energy are interpreted with the concept of
Skyrmion--quasiparticles \cite{SKKR93,FBCM94}.
These can be described as extended spin--textures containing a
number of spins which are flipped with respect to the preferential
direction set by the magnetic field.
Their number is determined by the competition between Zeeman
energy and Coulomb interaction of the electrons.
While the equilibrium of systems containing Skyrmions has been studied
in various ways -- both experimentally and theoretically, the Skyrmion's
dynamical properties are much less understood.
There are indirect measurements of electron spin resonance which use the
magnetoresisitivity to determine the spin--splitting of the Landau
levels \cite{SKW83,DKW88,DKSWP88}
and which also yield a line width \cite{DKW88}.

The first and fundamental question about the dynamics regards the
formation of the Skyrmion.
A Skyrmion anti--Skyrmion pair has only half the energy of a particle--hole
excitation, a spin exciton \cite{SKKR93,FBCM94};
nevertheless, it needs for its formation a mechanism which flips
the electron's spins, since a Skyrmion may contain a large number of
overturned spins.
The mechanism must be provided from outside the bare electron system
of Coulomb interaction and Zeeman energy, because the spin parallel
to the magnetic field is a constant of motion.
In this work, we wish to analyze the spin relaxation, originating
from such a mechanism, specifically for a system of 2-d interacting electrons
in a strong magnetic field.
As it will be shown below, the result is intriguing and non--trivial
for this system, already in the case of a uniform spin--density,
and quite different from the well--known Bloch equations \cite{Slichter}.
Thus, we will demonstrate our new approach studying the uniform case;
this has to be seen as the first step towards a solution of the more
general problem of the formation of the Skyrmion.

We consider in the following interacting electrons in two dimensions
moving in a strong magnetic field under the following conditions:
the electron density should be close to that of one filled spin--split
Landau sub--level, i.e.~the filling factor $\nu \simeq 1$,
and the Zeeman energy should be much less than the Coulomb energy.
Both conditions are relevant for the Skyrmion case.
The latter is realized in the experiment by a variation of the
$g$--factor with pressure \cite{MPPHEHP96}.
In three--dimensional (3-d) semiconductors, various mechanisms
of spin relaxation have been studied experimentally (e.g., in optical
measurements) and theoretically.
Among them are scattering by holes due to the exchange energy,
and scattering by phonons and impurities which becomes possible because
of the spin--orbit interaction; for a comparison of these different
mechanisms, see Ref.~\cite{APT83}.
In a system in strong magnetic field, the Zeeman energy has to be 
taken up by the scatterer in an elementary spin flip.
Therefore, we study here a general model of inelastic scattering
by phonons.
In deriving the kinetic equation which governs the relaxation of
the spin--density due to the scattering by phonons, the following
three assumptions are made:
(i) the 2-d electrons are described by a non--equilibrium distribution
which yields {\it uniform} expectation values for density and spin--density.
(ii) the phonons are in thermal equilibrium unperturbed by the electrons.
(iii) the electron--phonon coupling is so weak that it can be treated
in perturbation theory.
The appropriate technique for performing perturbation theory in
non--equilibrium systems was introduced by Keldysh \cite{K65,LL10P91}.
It is shown in Ref.~\cite{LL10P91} how to derive the kinetic equation
in the quasi--classical regime.
The condition for the applicability of the quasi--classical approximation
in the present context of the spin relaxation is given below.

How can phonons flip the spin ?
Consider the standard expression of the spin--orbit energy,
\be
 V_{so} = \frac{1}{2} \sig \cdot
         \frac{- e \hbar}{2 m^2 c^2} {\vec E} \times  {\vec p} \; ,
\label{Vso}
\ee
where $\frac{1}{2} \sig$ and $\vec p$ are spin and momentum
of the electron.
The electric field $\vec E$ has its origin in a piezoelectric
distortion of the lattice and it is linear in phonon operators.
Thus, the electrons are scattered and their spin is flipped by the
phonons of the 3-d lattice which contains the 2-d electron system
at $z=0$.
The orbital part of the electron's single particle states is confined
to the lowest Landau level.
We denote the projected spin--density of the electrons at (2-d) wave vector
$\vec q$ with ${\hat {\vec S}}({\vec q})$ \cite{BMV96}.
Then, we get for the electron--phonon part of the Hamiltonian
\be
  H_{e-ph} =  \sum_{{\vec Q},s}   {\hat {\vec S}}({\vec q})
  \left( \lam_{s}(-{\vec Q}) \; b^{}_{-{\vec Q},s}
         + \lam_{s}({\vec Q})^{*} \; b^{\dagger}_{{\vec Q},s} \right)
   \, . \label{Heph}
\ee
$b^{\dagger}_{{\vec Q},s}$ creates a phonon with (3-d) wave vector
${\vec Q} = ({\vec q},Q^z)$ and polarization $s$.
The coupling constant $\lam_{s}({\vec Q})$ can be derived
from (\ref{Vso}).
It containes the common electronic wave function in the z--direction.
With a relation between electric field and phonons as in the model of
Refs.~\cite{DMPT86,F91}, $\lam_{s}({\vec Q})$ is quadratic in
momentum. 
The considerations below are valid for arbitrary $\lam_{s}({\vec Q})$. 

Having specified the model, we will now study first the kinetic equation
for a strongly simplified case in which only two non--interacting electronic
states are kept.
In a second step, we will then come back to the full model of interacting
electrons in two Landau sub--levels coupled to a phonon bath.

\vspace{2ex}
{\it Two non--interacting states ---}
The Hamiltonian is given by
\bea
  H_{toy} =
   &-& \Delta (\au^{\dagger} \au - \ad^{\dagger} \ad)
      + \sum_{\vec q} \omega_{\vec q} \,
         \left(b^{\dagger}_{\vec q} b^{\,}_{\vec q} + \frac{1}{2}  \right)
          \nonumber \\
   &+& \sum_{\vec q} \lambda_{\vec q} \,
        \left( \ad^{\dagger} \au \; b^{}_{\vec q} +
               \au^{\dagger} \ad \; b^{\dagger}_{\vec q} \right)   \;,
\eea
where $c_{\uparrow,\downarrow}^{\dagger}$ ($b^{\dagger}_{\vec q}$)
are electron (phonon) creation operators.
Here, scattering from the $\uparrow$ state to the $\downarrow$ state occurs
under absorption of a phonon with an arbitrary wavevector $\vec q$,
and the same is true for the reverse process.
The kinetic equation for the time--dependent average occupation numbers
$n_{\uparrow,\downarrow}(t)$ can be derived with the aid of Fermi's
golden rule (here, time arguments are omitted):
\be
  \partial_t n_{\uparrow}  = - \partial_t n_{\downarrow} =
  \frac{1}{\tau} \left[ (1-n_{\uparrow}) n_{\downarrow} (1+N)
           - n_{\uparrow} (1-n_{\downarrow}) N \right]     \,.
\label{KEn}
\ee
Eq.\ (\ref{KEn}) is valid in the quasi--stationary limit $\Delta >> 1/\tau$,
and the relaxation time $\tau$ is given by
\be
 \frac{1}{\tau} = \sum_{\vec q} \lambda_{\vec q}^2 \;
          2 \pi \; \delta( 2 \Delta - \omega_{\vec q} ) \;.
\ee
$N = [ e^{\beta 2 \Delta} - 1 ]^{-1}$ is the occupation number of the only
phonon ($\omega_{\vec q} = 2 \Delta$) which can be effective under energy
conservation.
The density $\nu = n_{\uparrow}(t) + n_{\downarrow}(t)$ is fixed; thus it
is clear that Eq.~(\ref{KEn}) has two stationary solutions for
$n_{\uparrow}$.
Only one of these is physical, the other is unphysical ($n_{\uparrow}>1$).
The kinetic equation (\ref{KEn}) is best discussed in terms of the
depolarization, the deviation from the physical solution,
$\delta (t) = n_{\uparrow}(\infty) - n_{\uparrow}(t)$ :
\be
 \tau \; \partial_t \, \delta (t) =
   - \delta (t) \; [ \epsilon + \delta (t) ]    \;.
\label{KE}
\ee
The crucial parameter $\epsilon$ is the difference between the two
stationary solutions,
$\epsilon = \left[ 4 N (1+N) + (\nu -1)^2 \right]^{1/2}$.
It depends strongly on the temperature and the total density.
For zero temperature $1/\beta=0$, $\epsilon = |\nu -1|$.
For $\nu =1$, on the other hand, $\epsilon \sim 2 e^{- \beta \Delta}$
at low temperatures.
Now it is obvious that, and under which conditions, linearization in the
kinetic equation (\ref{KE}) can fail:
Eq.~(\ref{KE}) shows two regimes depending on the relative size
of $\epsilon$ and $\delta(t)$.
If $\delta (t)$ is smaller than $\epsilon$, then one can linearize
Eq.~(\ref{KE}) in $\delta (t)$, and $\delta (t)$ decays exponentially.
If $\delta (t)$ is larger than $\epsilon$, then Eq.~(\ref{KE}) becomes
quadratic in $\delta (t)$, and $\delta (t)$ decays with $1/t$ until
it becomes so small that it crosses over into the first regime.
This behavior is pictured in Fig.~1 for $\delta (t=0)=0.25$.
The solid line shows the depolarization for $\epsilon = 0.05$.
Initially, the curve follows that of the quadratic equation
($\epsilon =0$, dashed line), until it crosses over to exponential behavior.
This limiting exponential behavior is indicated as a dotted line.
We conclude that even the solution of the oversimplified model
$H_{toy}$ shows a non--trivial behavior, far from being
exponential {\it per se}.
It displays a strong dependence on temperature and density.
The asymptotic exponential decay of $\delta (t)$ can become
arbitrarily slow ($\epsilon \rightarrow 0$) as the temperature
approaches $0$ and $\nu \rightarrow 1$;
and in the limit, $\delta (t) \propto 1/t$.

\begin{figure}[thb]
\setlength{\epsfxsize}{0.46\columnwidth}
\centerline{\Bildeins}
Fig.\ 1: Depolarization $\delta (t)$ as a function of $t/\tau$
for an initial value of $\delta (0) = 0.25$;
solid line: $\epsilon = 0.05$, dashed line: $\epsilon = 0$.
\end{figure}

\vspace{2ex}
{\it Electron--electron interaction --- }
We now turn back to the case of interacting electrons in the two
lowest Landau sub--levels split by the Zeeman energy.
The electrons are scattered by phonons, see $H_{e-ph}$, Eq.~(\ref{Heph}).
The electronic part $H_e$ of the total Hamiltonian
$H_e + H_{ph} + H_{e-ph}$ is then
\be
H_e = - \Delta \sum_p (\aup^{\dagger} \aup^{} - \adp^{\dagger} \adp^{})
      + H_{Coul}  \;.
\ee
Here, $\aup^{\dagger}$ ($\adp^{\dagger}$) creates a Landau state with
linear momentum $p$ and spin $\uparrow$ ($\downarrow$).
The most important ingredient, which was missing in $H_{toy}$, is the
Coulomb interaction $H_{Coul}$ between the electrons, cf.~\cite{AB97}.
As stated in assumption (i) above, the expectation values of density
and spin--density of the electrons are presumed to be uniform.
The density is conserved and its value is given by the filling factor
$\nu \sim 1$.
We wish to study the uniform spin--density,
${\vec S}(t) = <{\hat {\vec S}}({\vec q} =0)> $,
\be
 {\vec S}(t) =
     \frac{1}{N_\Phi} \sum_p
  <  (\aup^{\dagger} , \adp^{\dagger} )(t) \; \frac{1}{2} \sig \;
      {\aup \choose \adp }(t) >     \;.
\ee
$N_\Phi$ denotes the number of states in a Landau sub--level;
if all $\uparrow$ states are populated and all $\downarrow$ states are
empty, ${\vec S}(t) = \frac{1}{2} \hat e_z$.
The derivation of the kinetic equation for ${\vec S}(t)$ can be performed
with the method of Keldysh \cite{K65}.
The kinetic equation contains a precession term,
which is already present without any phonon scattering,
and the main term resulting from the collisions with the phonons.
In leading (second) order perturbation theory in the electron--phonon
coupling, the collision integral becomes
\bea
 &&\partial_t  S^m(t) |_{coll} =  - i \int^t_{-\infty} dt'
  \sum_{\vec Q} \sum_{j,k,l}
  \epsilon^{mjk}   \nonumber \\
 && \left[ {\cal C}^{kl}({\vec q};t,t') {\cal D}^{jl}({\vec Q};t-t')
         - {\cal C}^{lk}({\vec q};t',t) {\cal D}^{lj}({\vec Q};t'-t) \right] .
\nonumber \\  &&    \label{coll}
\eea
Only times $t'<t$ contribute due to causality.
The function $\cal C$ is the dynamical spin--spin correlation function,
\be
 {\cal C}^{kl}({\vec q};t,t') =  \frac{1}{N_\Phi}
< {\hat S}^k({\vec q},t) \; {\hat S}^l({-\vec q},t') > \; .
\ee
The spin--density couples in $H_{e-ph}$ to the phonons via
\be
  {\hat {\vec \Phi}}({\vec Q}) = \sum_s
  \left( \lam_{s}(-{\vec Q}) \; b^{}_{-{\vec Q},s}
       + \lam_{s}({\vec Q})^{*} \; b^{\dagger}_{{\vec Q},s} \right)
\; ;
\ee
thus, the phonon expectation value $\cal D$ in (\ref{coll}) is given by
\be
{\cal D}^{jl}({\vec Q};t-t')
 = <{\hat \Phi}^j({\vec Q},t) \; {\hat \Phi}^l({-\vec Q},t')  > \; .
\ee
$\cal D$ is easily calculated with the equilibrium Hamiltonian $H_{ph}$
of the phonons.
$\epsilon^{ijk}$ is the antisymmetric tensor.
Its origin is the time derivative in (\ref{coll}) which leads to a commutator
of $\hat {\vec S}$ with the spin--density in $H_{e-ph}$.

It is apparent now that {\it collective modes} are responsible for
the relaxation process.
The question is whether one can express the two--particle (four--fermion)
correlation function of the interacting system, $\cal C$, again by 
${\vec S}(t)$ and thus derive a closed equation for ${\vec S}(t)$.
Fortunately, that turns out to be possible in our case:
The time dependence of $\cal C$ is determined by particle--hole excitations
(e.g., $\hat S^+ = \hat S^x + i  \hat S^y \sim \au^{\dagger} \ad$).
For $H_e$, the particle--hole excitations above the ground state are
spin--excitons and their dispersion $E_{ex}(q)$ is rigorously
known \cite{BIE81,KH84} at $\nu=1$,
\be
 E_{ex}(q) = 2 \Delta + \frac{e^2}{\kappa l_B} \sqrt{\frac{\pi}{2} }
  \left[ 1 - e^{-\frac{q^2 l_B^2}{4}} I_0(\frac{q^2 l_B^2}{4}) \right] \;,
\ee
$\kappa$ is the dielectric constant, $l_B$ is the magnetic length,
and $I_0(x)$ denotes the Bessel function.
We now approximate, in the present case of a non--equilibrium state
and $\nu \sim 1$, the time dependence of $\cal C$ with the above dispersion
of the spin--exciton, $E_{ex}(q)$.
Thus we neglect (i) the difference between the real energy of a single
spin--exciton in a non--equilibrium state and $E_{ex}(q)$,
and (ii) interactions between the spin--excitations at $\nu \neq 1$.
This yields, e.g.,
\be
  {\cal C}^{k+}({\vec q};t,t') \simeq \; e^{i E_{ex}(q) (t-t') } \;
  {\cal C}^{k+}({\vec q};t,t)  \;.
\ee
The remaining equal--time correlations are then calculated in the
Hartree--Fock approximation.
This is the weak--scattering approximation in which the effect of the
phonons is included in the lowest order in the state, but neglected in the
time dependence \cite{LL10P91}.
Since the characteristic energy for the time dependence is the
Zeeman energy $\Delta$, that corresponds to the condition $\Delta >> 1/\tau$,
where the scattering time $\tau$ is defined below in Eq.\ (\ref{tau}).
The two approximations above regarding the exciton energy demand that
the electronic temperature must be small compared to the Zeeman energy
$\Delta$ and also, the filling factor must be such that $| \nu -1 | << 1$.
Now, collecting all terms, we get the resulting kinetic equations
for the components of  ${\vec S}(t)$  ($S^+ = S^x + i S^y$;
details of the calculation are deferred to a forthcoming publication):
\bea
 && \partial_t S^z(t) =
   \frac{1}{\tau}  \left[ 1- \frac{\nu}{2} - S^z(t) \right]
                    \left[ \frac{\nu}{2} - S^z(t) \right]
    - \frac{2}{\tau} {\bar N} S^z(t)   \nonumber \\
 &&  \label{resKE} \\
 && ( \partial_t  + i 2 \Delta ) \; S^+(t) =
   - \frac{1}{\tau} \left[ \frac{1}{2}- S^z(t)+ {\bar N} \right] \; S^+(t) \;.
 \nonumber 
\eea
The relaxation time $\tau$ and the average phonon number $\bar N$ are defined
by the two components of the following equation
(here, $N(\omega) = [ e^{\beta  \omega} - 1 ]^{-1}$)
\be
 \left[ { 1/\tau \atop {\bar N}/\tau } \right]
  =  \sum_{{\vec Q},s \atop j=x,y}
     \frac{\pi}{2} |\lambda^{j}_{s}({\vec Q})|^2  \;
      \delta( E_{ex}(q) - \omega_{{\vec Q},s} )
 \left[ { 1 \atop N(\omega_{{\vec Q},s}) } \right] .
\label{tau}
\ee
It is quite instructive to formulate again the kinetic equation in terms
of the depolarization vector $\delt(t)$.
We use $\delta^z(t) = S^z(\infty) - S^z(t)$ and we split the precession
term off $S^+(t)$ by redefining $\delta^+(t) = \exp(i 2 \Delta t) S^+(t)$.
Then, the result is
\bea
 \tau \; \partial_t \, \delta^z(t) &=& - \delta^z(t)  \;
                           [ \epsilon + \delta^z(t) ]  \nonumber \\
 \tau \; \partial_t \, \delta^+(t) &=& - \delta^+(t)  \;
                           [ \frac{\epsilon}{2} + \delta^z(t) ]  \;.
\label{KEs}
\eea
Here, the parameter $\epsilon$ corresponding to the one used above in the
case of the two--state model is defined as
$\epsilon = \left[ 4 {\bar N} (1+{\bar N}) + (\nu -1)^2 \right]^{1/2}$.
Obviously, Eqs.~(\ref{KEs}) are quite different from the standard
Bloch equations \cite{Slichter}.
They are non--linear and it is seen that the relaxation of the transverse
component $\delta^+(t)$ depends on the other component.
Since the first of the two kinetic equations (\ref{KEs}) is identical to
(\ref{KE}), the same discussion applies.
Observing $\tau \partial_t \ln [\delta^z(t)/\delta^+(t)] = -\epsilon /2$,
Eqs.\ (\ref{KEs}) can be explicitely solved with the result
\be
\frac{\delta^z(t)}{\delta^z(0)}  =
  e^{- \epsilon t/ (2 \tau)}  \frac{\delta^+(t)}{\delta^+(0)}  =
      \frac{e^{- \epsilon t/ \tau}}
        {1 + \delta^z(0) (1- e^{- \epsilon t/ \tau}) / \epsilon }  \; .
\ee
For any finite value of the parameter $\epsilon$, the leading asymptotic
behavior yields a ratio $T_2 = 2 T_1$ of the relaxation times.
On the other hand, for zero temperature and $\nu =1$, all normalized
components of the depolarization follow {\it the same function}
which is a power law $\propto t^{-1}$.

The relaxation time $\tau$ is determined by those phonons, whose energy and
in--plane momentum match the energy and the momentum of the spin--excitons,
see Eq.~(\ref{tau}).
Using the model of D'yakonov and Perel' (c.f.\ Ref.\ \cite{DMPT86,F91}),
we calculate the electron phonon coupling parameter
$\lambda^j_s({\vec Q})$ and perform the summations and integrations
in (\ref{tau}) with the following result in the limit of $\Delta =0$
\be
  \frac{1}{\tau} = \frac{1}{8} \sqrt{\frac{2}{\pi}} \;
    \left( \frac{v}{s} \right)^2
    \frac{\hbar}{\rho l_B^5} \; x_0 \, e^{-4x_0} \, W_{1,1}(8x_0) \;.
\label{restau}
\ee
Here, $v$ parametrizes the electron phonon coupling, $s$ is the phonon
velocity, $\rho$ the 3-d density, and $W_{1,1}(x)$ the Whittaker function.
The parameter $x_0$ is given by
$x_0 = (\hbar s / (\epsilon_c l_B))^2$, where $\epsilon_c$ is the Coulomb
energy.
An estimate with characteristic values for electrons in GaAs at a magnetic
field of 10 T yields $\tau \sim 10^{-10}sec$.
This order of magnitude is much smaller than the estimate 
in Ref.~\cite{F91} and in agreement with experimental
observation \cite{DKW88}.

We have studied the spin relaxation in a quantum Hall system for
filling factors $\nu \simeq 1$.
The resulting kinetic equation (\ref{resKE}) is non--linear
and quite different from a conventional Bloch equation.
This difference is most pronounced for $\nu \rightarrow 1$ and
$T \rightarrow 0$;
there, a linearization of the kinetic equation fails,
since the linear term vanishes at $\nu =1$, $T=0$.
The relevant time $\tau$ (\ref{tau},\ref{restau})
in the kinetic equation is determined dominantly
by the Coulomb interaction between the electrons. ---

We wish to thank R. Haug for a careful reading of this manuscript. 
Yu.~B. thanks the PTB and W.~A. thanks the Landau Institute, 
respectively, for their hospitality. 
We acknowledge support 
by the U.S. Civilian Research and Development Foundation
under Award \#RP1-273, and RFFI Grant N 97-02-16042 (Yu.~B.)
and by the Deutsche Forschungsgemeinschaft, Ap 47/1-2 (W.A.).

\end{multicols}

\begin{thebibliography}{10}

\bibitem{NHKW88}
R.~J. Nicholas, R.~J. Haug, K. von Klitzing, and G. Weimann, Phys. Rev. B {\bf
  37},  1294  (1988).

\bibitem{UNHF90}
A. Usher, R.~J. Nicholas, J.~J. Harris, and C.~T. Foxon, Phys. Rev. B {\bf 41},
   1129  (1990).

\bibitem{BDPWT95}
S.~E. Barrett {\it et~al.}, Phys. Rev. Lett. {\bf 74},  5112  (1995).

\bibitem{SEPW95}
A. Schmeller, J.~P. Eisenstein, L.~N. Pfeiffer, and K.~W. West, Phys. Rev.
  Lett. {\bf 75},  4290  (1995).

\bibitem{AGB96}
E.~H. Aifer, B.~B. Goldberg, and D.~A. Broido, Phys. Rev. Lett. {\bf 76},  680
  (1996).

\bibitem{MPPHEHP96}
D.~K. Maude {\it et~al.}, Phys. Rev. Lett. {\bf 77},  4604  (1996).

\bibitem{KKE97}
I.~V. Kukushkin, K. von Klitzing, and K. Eberl, Phys. Rev. B {\bf 55},  10607
  (1997).

\bibitem{LNMUPHF97}
D.~R. Leadley {\it et~al.}, Phys. Rev. Lett. {\bf 79},  4246  (1997).

\bibitem{SKKR93}
S.~L. Sondhi, A. Karlhede, S.~A. Kivelson, and E.~H. Rezayi, Phys. Rev. B {\bf
  47},  16419  (1993).

\bibitem{FBCM94}
H.~A. Fertig, L. Brey, R. C\^ot\'e, and A.~H. MacDonald, Phys. Rev. B {\bf 50},
   11018  (1994).

\bibitem{SKW83}
D. Stein, K. von Klitzing, and G. Weimann, Phys. Rev. Lett. {\bf 51},  130
  (1983).

\bibitem{DKW88}
M. Dobers, K. von Klitzing, and G. Weimann, Phys. Rev. B {\bf 38},  5453
  (1988).

\bibitem{DKSWP88}
M. Dobers {\it et~al.}, Phys. Rev. Lett. {\bf 61},  1650  (1988).

\bibitem{Slichter}
C.~P. Slichter, {\em Principles of Magnetic Resonance}, {\em Springer Series in
  Solid-State Sciences, Vol. 1} (Springer--Verlag, Berlin Heidelberg New York,
  1980).

\bibitem{APT83}
A.~G. Aronov, G.~E. Pikus, and A.~N. Titkov, Sov. Phys. JETP {\bf 57},  680
  (1983).

\bibitem{K65}
L.~V. Keldysh, Sov. Phys. JETP {\bf 20},  1018  (1965).

\bibitem{LL10P91}
L.~D. Landau and E.~M. Lifshitz,  in {\em Physical Kinetics}, Vol.~X of {\em
  Course of Theoretical Physics} (Pergamon, Oxford, 1981), Chap.~\S 91.

\bibitem{BMV96}
Yu.~A. Bychkov, T. Maniv, and I.~D. Vagner, Phys. Rev. B {\bf 53},  10148
  (1996).

\bibitem{DMPT86}
M.~I. D'yakonov, V.~A. Marushchak, V.~I. Perel', and A.~N. Titkov, Sov. Phys.
  JETP {\bf 63},  655  (1986).

\bibitem{F91}
D.~M. Frenkel, Phys. Rev. B {\bf 43},  14228  (1991).

\bibitem{AB97}
W. Apel and Yu.~A. Bychkov, Phys. Rev. Lett. {\bf 78},  2188  (1997).

\bibitem{BIE81}
Yu.~A. Bychkov, S.~V. Iordanskii, and G.~M. \'{E}liashberg, JETP Lett. {\bf 33},
   143  (1981).

\bibitem{KH84}
C. Kallin and B.~I. Halperin, Phys. Rev. B {\bf 30},  5655  (1984).

\end{thebibliography}
\end{document}